\documentclass[aps,pra,reprint,superscriptaddress]{revtex4-1}

\usepackage{amsmath}
\usepackage{graphicx}
\usepackage{dcolumn}
\usepackage{bm}

\def\ket#1{\mathinner{|{#1}\rangle}}

\newcommand{\etal}{{\it et al.\ }}

\begin{document}

\title{Observation of the nuclear magnetic octupole moment of $^{173}$Yb from precise measurements of hyperfine structure in the ${^3P}_2$ state}
 \author{Alok K. Singh}
 \affiliation{Department of Physics, Indian Institute of
 Science, Bangalore 560\,012, India}
 \author{D. Angom}
 \affiliation{Physical Research Laboratory, Navarangpura 380\,009, India}
 \author{Vasant Natarajan}
 \affiliation{Department of Physics, Indian Institute of
 Science, Bangalore 560\,012, India}
 \email{vasant@physics.iisc.ernet.in}
 \homepage{www.physics.iisc.ernet.in/~vasant}

\begin{abstract}
We measure hyperfine structure in the metastable ${^3P}_2$ state of $^{173}$Yb and extract the nuclear magnetic octupole moment. We populate the state using dipole-allowed transitions through the ${^3P}_1$ and ${^3S}_1$ states. We measure frequencies of hyperfine transitions of the ${^3P}_2 \rightarrow {^3S}_1$ line at 770 nm using a Rb-stabilized ring cavity resonator with a precision of 200 kHz. Second-order corrections due to perturbations from the nearby ${^3P}_1$ and ${^1P}_1$ states are below 30 kHz. We obtain the hyperfine coefficients as: $A=-742.11(2)$ MHz, $B=1339.2(2)$ MHz, which represent two orders-of-magnitude improvement in precision, and $C=0.54(2)$ MHz. From atomic structure calculations, we obtain the nuclear moments: quadrupole $Q=2.46(12)$ b and octupole $\Omega=-34.4(21)$ b\,$\times \mu_N$.
\end{abstract}


\maketitle

\section{Introduction}
Observation of the nuclear magnetic octupole moment and its influence on the hyperfine structure of atoms has remained largely unexplored because of its weaker effect compared to the leading magnetic dipole and electric quadrupole moments. The most recent measurements have been on one-electron systems---the $5D_{3/2}$ state of $^{137}$Ba$^+$ ion \cite{LCC12}, and the $6P_{3/2}$ state of $^{133}$Cs atom \cite{DAN05,GDT03}. But the sub-kHz value of the hyperfine coefficient required similar precision in measuring the intervals. By contrast, the long-lived ${^3P}_2$ state in two-electron atoms such as Yb and several alkaline-earth metals, with its large angular momentum, is a more sensitive probe to observe this moment \cite{BDJ08}. Here, we report precise measurement of hyperfine structure in the ${^3P}_2$ state of $^{173}$Yb, and observation of the magnetic octupole moment using calculations in two-electron atoms. In a previous measurement of hyperfine structure in two isotopes of Eu, $^{151}$Eu and $^{153}$Eu, the author was only able to extract the isotopic ratio of the octupole moment \cite{CHI91}. Precise measurement of hyperfine structure is also motivated by the fact that its comparison to theoretical calculations plays an important role in validating the atomic wavefunctions used in the calculations. In this regard, Yb is an important atom because of its proposed use in search for a permanent electric dipole moment (EDM) \cite{NAT05,PRS10}, where comparison to calculation \cite{MAA11} forms a vital tool in searching for new physics beyond the Standard Model. The ${^3P}_2$ state in Yb has potential applications in more sensitive EDM searches \cite{PRP10}, and ultra-sensitive magnetometry \cite{YUT08}. The presence of nuclear octupole deformation has been shown to lead to an enhanced collective EDM that can significantly exceed single-particle moments \cite{PGF11}.

Measurements on upper levels is an experimental challenge because these levels are not directly populated. We have earlier solved this problem by using dipole-allowed transitions to pump atoms into the metastable ${^3P}_2$ state of Yb \cite{PRP10}. In this work, we use the same method to populate this state, and then measure the absolute frequencies of various hyperfine transitions on the ${^3P}_2 \rightarrow {^3S}_1$ line. We measure the frequencies with our well-developed technique of using a Rb-stabilized ring-cavity resonator \cite{BDN03,BRD03,DBB06,PSK09}. We obtain the hyperfine structure coefficients $A$ (magnetic dipole) and $B$ (electric quadrupole) with two orders-of-magnitude better precision than previous values, and a 4\% measurement of the magnetic octupole coefficient $C$. We take into account second-order corrections due to perturbations from the nearby ${^3P}_1$ and ${^1P}_1$ states, and find that they make negligible contribution. Using relativistic coupled-cluster (RCC) calculations, we obtain the ratios of these hyperfine coefficients to the relevant nuclear moments, namely $B/Q$ and $C/\Omega$. From this, we obtain the quadrupole moment $Q=2.46(12)$ b and the octupole moment $\Omega=-34.4(21)$ b\,$\times \mu_N$.

\section{Experimental details}
\begin{figure}
\centering{\resizebox{0.95\columnwidth}{!}{\includegraphics{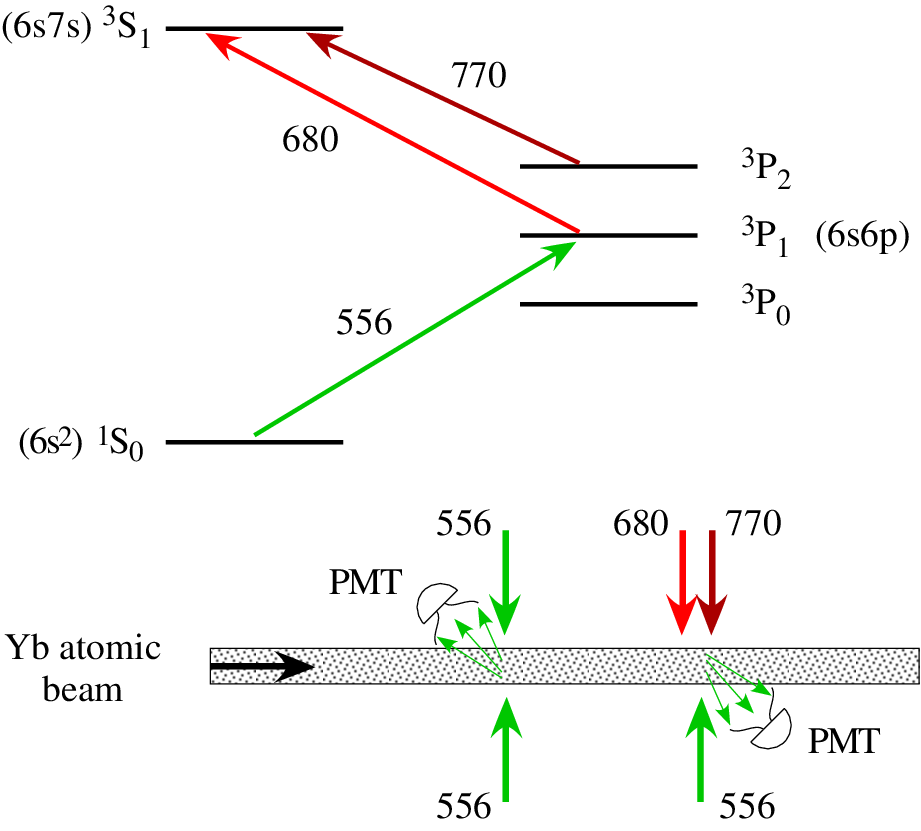}}}
\caption{(Color online) On top are the relevant low-lying energy levels of Yb showing the wavelength of each transition in nm. Below is a schematic of the spectroscopy chamber for optically pumping atoms into the ${^3P}_2$ state, and measuring the ${^3P}_2 \rightarrow {^3S}_1$ transition at 770 nm.}
 \label{specchamber}
\end{figure}

The relevant low-lying energy levels of Yb are shown in Fig.\ \ref{specchamber}. The ground state is ${^1S}_0$ and therefore has no hyperfine structure. But the upper states with $J \neq 0$ have hyperfine levels for the odd isotope $^{173}$Yb, determined by the nuclear spin $I=5/2$. The various transitions are accessed in the spectroscopy chamber, shown schematically in the bottom of the figure. It consists of a vacuum chamber with several optical access points maintained at a pressure below $10^{-8}$~mbar with an ion pump. The Yb atomic beam is generated by resistive heating (to about 400$^\circ$C) of an unenriched source. The different laser beams are sent perpendicular to the atomic beam at different points, and the green fluorescence (at 556~nm) is collected by two photomultiplier tubes (PMT, Hamamatsu R928). The 556~nm beam, driving the ${^1S}_0 \rightarrow {^3P}_1$ transition, is produced by doubling the output of a fiber laser operating at 1111~nm (Koheras Boostik Y10). The output power of the fiber laser is 0.5~W with a linewidth of 70~kHz. It is doubled in an external cavity doubler to give a total power of 65~mW. Part of this beam is split into two and sent across the atomic beam in counter-propagating directions, perpendicular to the atomic beam. The laser is locked to the peak center by frequency modulation at 20~kHz and lock-in detection to generate the error signal.

The lasers at 680~nm (driving the ${^3P}_1 \rightarrow {^3S}_1$ transition) and 770~nm (driving the ${^3P}_2 \rightarrow {^3S}_1$ transition) are home-built diode laser systems \cite{BRW01}. They are frequency stabilized using grating feedback to give linewidths of order 1~MHz. The 680 beam counterpropagates with the locked 556 beam a few cm downstream from the 556 fluorescence point. {\em It optically pumps atoms into the metastable ${^3P}_0$ and ${^3P}_2$ states}. This reduces population in the ground state and the green fluorescence spectrum when scanning the 680 beam shows a negative peak. The injection current into this laser is
frequency modulated (at 15~kHz) and it is locked to the
peak center. The 770 beam, which is a further 2~mm
downstream and overlaps with the same 556 beam, causes the
fluorescence to recover as it repumps atoms from the
${^3P}_2$ state back into the ground state, from where they can reabsorb the 556 beam. Thus, {\em the 770 beam measures the population in the ${^3P}_2$ state} and the fluorescence spectrum shows a positive peak. The measurements are done with this laser locked to a particular hyperfine transition. For this, the laser is current modulated at a frequency of 10 kHz, which is different from that of the 680 laser, and the same PMT signal is demodulated at this frequency to generate the error signal.

Representative fluorescence spectra for the even isotope $^{174}$Yb obtained by scanning each of the three lasers have been shown in our previous work \cite{PRP10}. For the odd isotope $^{173}$Yb used in this work, the number of hyperfine levels and the electric dipole selection rules shows that there are 9 possible transitions at 770 nm, of which 6 are used for the measurement. The spectra for the measured transitions appear with different signal-to-noise ratios (SNR) depending on the individual transition strengths. The SNR in all cases is good enough to get a strong error signal to lock the laser.

Our frequency-measurement technique using the Rb-stabilized ring-cavity resonator has been described extensively in earlier work \cite{BDN03,DBB06}, and is reviewed here for completeness. It relies on the fact that the frequency of the 780-nm $D_2$ line in $^{87}$Rb ($5S_{1/2} \leftrightarrow 5P_{3/2}$ transition) is known with an accuracy of 6 kHz \cite{YSJ96}. A diode laser locked to a particular hyperfine transition of this line is used as a frequency reference. A second laser is in turn locked to the unknown transition. The two lasers are coupled into an {\em evacuated} ring-cavity resonator. An acousto-optic modulator (AOM) placed in the path of one of the two lasers is used to produce a small frequency offset (of order 100 MHz) so that the cavity is in simultaneous resonance with both laser frequencies. Thus the ratio of the unknown frequency to the reference frequency is just a ratio of two integers (i.e.\ the respective cavity mode numbers) combined with the AOM offset. The procedure to determine the unique mode-number combination for the two lasers has been described earlier \cite{DBB06}. We have used this technique to measure frequencies of transitions in the range of 670 nm to 895 nm, i.e.\ $\pm 100$ nm away from the reference wavelength.

\section{Error analysis}
Our technique is particularly well-suited to the measurement of frequency differences (as used in the measurement of hyperfine intervals). This is because several sources of systematic errors cancel when taking the difference. Systematic errors can be classified under three categories: errors related to locking the reference laser to a particular hyperfine transition, errors related to locking the unknown laser on the Yb line, and errors due to the measurement cavity. Errors related to the reference laser are the same for all measurements and therefore cancel in the difference. Similarly, errors due to dispersion inside the cavity and at the cavity (multilayer dielectric) mirrors do not affect the difference. Effects of geometric misalignment into the cavity result in the excitation of higher-order modes, which can pull the lock point of the fundamental mode. To the extent that the misalignment is the same for all measurements, this will again not affect the difference. Thus all the sources of systematic error in the
difference frequency are related to how well we can lock
the 770-nm laser to a particular Yb transition and how
well we can lock the cavity to the laser, as discussed below.

Collisional shifts of the transition frequency are minimal
because we use an atomic beam. A systematic Doppler shift of the peak center will occur if the angle between the atomic beam and the laser beam is not exactly 90$^\circ$. However,  this will cause a negligible error when taking the difference---for example, even a large misalignment angle of 10 mrad will result in an error of only 2 kHz when measuring a 5-GHz interval. Another potential source of systematic shift in the peak position is due to lineshape asymmetry that might occur because of selective driving into Zeeman sublevels (shifted in the presence of stray magnetic fields) or radiation pressure. The first effect is minimized by using linearly polarized light so that the Zeeman sublevels are excited equally about line center. The maximum Zeeman shift among all transitions in the presence of a stray field of 10 mG is about 70 kHz. By studying the symmetry of the fluorescence lineshape, we conclude that this error is smaller than 70 kHz. Thus, the main source of error in our measurement is determined by how well we lock to peak center---split the line---given the linewidth of about 40 MHz and the SNR. In our recent work on measuring isotope shifts in the 556-nm line of Yb \cite{PSK09}, the linewidth was about 6 MHz and the locking error was 30 kHz. With the 6 times larger linewidth here, we estimate the error to be 200 kHz (or equal to splitting the line by about 1 part in 200), which is larger than all other sources of error. As we will see below, we have a good experimental handle on this estimate by measuring the same transition with two different transitions to lock the reference laser---one from the $F=1 \rightarrow F'$ hyperfine manifold and the other from the $F=2 \rightarrow F'$ manifold.

\section{Results and discussion}
We have measured the frequencies of various hyperfine transitions by locking the 770-nm laser to different peaks. For each measurement, the time constant in the frequency counter (with a timebase stability of $10^{-8}$) was set to 10 s. Then a set of 40 independent measurements was made and the average determined. For each transition, the frequency was measured with the reference laser on either a $F=1 \rightarrow F'$ transition or a $F=2 \rightarrow F'$ transition. The resulting change in the frequency is on the order of 6.5 GHz, and is known with $<10$ kHz precision from hyperfine measurements in $^{87}$Rb \cite{YSJ96}.

The measured frequencies are shown in Table \ref{freqs}. The error in each value is 0.2 MHz. In order to highlight the above fact that many sources of error cancel in the difference, we show the frequencies as offset from the first value. We have verified that if we make an error in the mode number and change it by $\pm 1$, the absolute frequency of all transitions changes by about 12 MHz, but there is {\em negligible} change in the offsets listed in the table. Our error estimate is reasonable because the two values for each transition, which implies a different set of cavity mode numbers and complete re-optimization of all the feedback loops, overlap quite well. In addition, the maximum standard deviation in each set of 40 measurements (from which the average is determined) is only 170~kHz. One of the self-consistency checks that we can perform on the error bar is the $\{ 7/2-5/2 \}$ hyperfine interval in the ${^3P}_2$ state. It can be evaluated in two ways: $(7/2 \rightarrow 5/2) - (5/2 \rightarrow 5/2)$ and $(7/2 \rightarrow 7/2) - (5/2 \rightarrow 7/2)$. The value of 0.22 MHz is consistent with zero when we take into account the error in each value of 0.2 MHz.

\begin{table*}
\caption{Measured frequencies for various hyperfine transitions of the ${^3P}_2 \rightarrow {^3S}_1$ line with the reference laser locked to either an $F=1 \rightarrow F'$ or an $F=2 \rightarrow F'$ transition. The values are given as offset from the first transition. The numbers in brackets are $1\sigma$ errors in the last digit.}
\label{freqs}
\begin{ruledtabular}
\begin{tabular}{ccccccc}
Ref.\ Laser & \multicolumn{6}{c}{Frequency (MHz)} \\
\cline{2-7}
lock point & $3/2 \rightarrow 5/2$ & $5/2 \rightarrow 5/2$ & $7/2 \rightarrow 5/2$ & $5/2 \rightarrow 7/2$ & $7/2 \rightarrow 7/2$ & $9/2 \rightarrow 7/2$ \\
\hline
$1 \rightarrow F'$ & 0 & 2523.43(20) & 5360.52(20) & $-4075.69(20)$ & $-1238.38(20)$  & 1193.92(20)  \\
$2 \rightarrow F'$ & 0.21(20) & 2523.56(20) & 5360.60(20) & $-4075.65(20)$ & $-1238.46(20)$ & 1193.97(20) \\
\end{tabular}
\end{ruledtabular}
\end{table*}

\subsection{Extracting the hyperfine coefficients}
In order to fit the measured intervals to the hyperfine coefficients, we first need to know the energy shift of an $F$ level due to the hyperfine interaction. From perturbation theory, the shift is given by
\begin{equation}
W_F=W_F^{(1)}+W_F^{(2)} \, ,
\label{shift}
\end{equation}
where the first-order shift arises due to the various nuclear moments. It has the progressively weaker magnetic-dipole ($A$), electric-quadrupole ($B$), and magnetic-octupole ($C$) terms. The prefactors multiplying these coefficients for each $F$ level in the ${^3P}_2$ state are well-known functions of $I$, $J$, and $F$ \cite{GDT03}, and are listed in Table \ref{factors}. The $A$, $B$, and $C$ coefficients are related to the corresponding nuclear moments: $\mu_I$, $Q$, and $\Omega$, respectively. In particular, $B = 2Q \langle T^e_2\rangle$, where $\langle T^e_2\rangle$ is the matrix element of the quadrupole field operator with the atomic state; and $C = -\Omega \langle T^e_3\rangle$, where $\langle T^e_3\rangle$ is the matrix element of the octupole field operator with the atomic state. Therefore, these matrix elements can be calculated if we have the wavefunction describing the ${^3P}_2$ state. For this, we use the Fock space RCC theory \cite{MAA11}. The atomic state is then defined as $e^T(1+S) \ket{\Phi_0}$, where $T$ and $S$ are the core and valence cluster operators and $\ket{\Phi_0}$ is the closed-shell reference state. For the RCC calculations, we use the no-virtual-pair Dirac-Coulomb Hamiltonian defined in our earlier work in Ref.\ \cite{MAA11}. It includes the nuclear Coulomb potential and the electron-electron Coulomb interactions.

\begin{table}
\caption{Prefactors multiplying the hyperfine coefficients of the first-order shift for each $F$ level in the ${^3P}_2$ state of $^{173}$Yb. The last row is the second-order correction to the shift in MHz.}
\label{factors}
\begin{ruledtabular}
\begin{tabular}{cccccc}
$F$ & 1/2 & 3/2 & 5/2 & 7/2 & 9/2 \\
\hline
$W_A^{(1)}$ & $-7$ & $-11/2$ & $-3$ & 1/2 & 5 \\
$W_B^{(1)}$ & 7/10 & 1/4 & $-1/4$ & $-17/40$ & 1/4 \\
$W_C^{(1)}$ & $-42/5$ & 12/5 & 27/5 & $-22/5$ & 1 \\
$W_F^{(2)}$ & 0 & $-0.05$ & 0.02 & 0.20 & 0 \\
\end{tabular}
\end{ruledtabular}
\end{table}

The second-order shift in Eq.~(\ref{shift}) arises mainly because of perturbations from the nearby ${^3P}_1$ and ${^1P}_1$ states \cite{LMN62}. If we limit ourselves to the dipole-dipole and dipole-quadrupole interactions, then the correction can be expressed in terms of the parameters $\eta$ and $\zeta$ defined in Ref.\ \cite{BDJ08}. The calculation involves summation over a complete set of intermediate states, and is done in the same way as described above \cite{MAA11}. Due to the near degeneracy of the ${^3P}_1$ state, it is the most-dominant intermediate state. The calculated second-order corrections are listed in the last line of Table \ref{factors}. We have also verified that these corrections are reasonable---we get similar values using an alternate method described in Ref.\ \cite{LMN62}, which uses measured hyperfine coefficients of the nearby states to estimate the values.

The procedure to extract the coefficients is now straightforward. The shift of each level in terms of $A$, $B$, $C$, and the second-order correction is known. This gives a set of 6 equations corresponding to the 6 measured transition frequencies. We then do a least-squares fit to these equations with the hyperfine coefficients as fit parameters. In order to see the relative importance of $C$ and the second-order correction, we have calculated the coefficients with and without these parameters. The results are shown in Table \ref{consts}. The difference from not including the second-order corrections is negligible, which is to be expected since the corrections are much smaller than our overall error bars. However, not including $C$ makes a large difference: of $15 \sigma$ (combined) in $A$ and of $11 \sigma$ (combined) in $B$. This shows that there is compelling evidence for $C$ from our measurement, and the extracted value of $0.54(2)$~MHz demonstrates that its value lies between 0.48 and 0.60 MHz with nearly 100\% confidence \footnote{The nuclear electric hexadecapole coefficient $D$ is expected to be negligible, but even a value of 3 kHz makes no significant difference to the $A,B,C$ coefficients.}.

\begin{table}
\caption{Calculated values of the hyperfine coefficients for the ${^3P}_2$ state in $^{173}$Yb. The second row is calculated without including the second-order correction and the third row is calculated without the magnetic octupole coefficient $C$. All values in MHz.}
\label{consts}
\begin{ruledtabular}
\begin{tabular}{lcccc}
& \multicolumn{1}{c}{$A$ } & \multicolumn{1}{c}{$B$ } & \multicolumn{1}{c}{$C$} & Reduced $\chi^2$ \\
\hline
Corrected & $-742.11(2)$ & 1339.2(2) & 0.54(2) & 0.23 \\
Uncorrected & $-742.11(2)$ & 1339.4(2) & 0.55(2) & 0.23 \\
Without $C$ & $-742.53(2)$ & 1342.5(2) & \multicolumn{1}{c}{--} & 118
\end{tabular}
\end{ruledtabular}
\end{table}

Our values are compared to previous values in Table \ref{comp}. The most recent experimental value is a 1992 measurement by Maier \etal \cite{MKB91}. The two sets of values are consistent but our error bars are about 100 times smaller. An earlier 1962 measurement by Ross and Murakawa \cite{ROM62} gives values in cm$^{-1}$ with no error bars. Their values appear close but we cannot judge the overlap without errors. Not surprisingly, none of these values are sensitive to the octupole coefficient $C$. There is also a theoretical calculation in 1999 by Porsev \etal \cite{PRK99}, which yields values within a couple of MHz (less than 0.5\%) of our values.

\begin{table}
\caption{Comparison of hyperfine coefficients for the ${^3P}_2$ state in $^{173}$Yb from this work to previous values.}
\label{comp}
\begin{ruledtabular}
\begin{tabular}{llcc}
\multicolumn{1}{c}{$A$ } & \multicolumn{1}{c}{$B$ } & \multicolumn{1}{c}{$C$ } & Reference \\
\hline
$-742.11(2)$ & 1339.2(2) & 0.54(2) & This Work\\
$-742(5)$ & 1342(38) & \multicolumn{1}{c}{--} & \cite{MKB91}\\
$-738$ & 1310 & \multicolumn{1}{c}{--} & \cite{ROM62}\footnotemark[1] \\
$-745$ & 1335 & \multicolumn{1}{c}{--} & Theory \cite{PRK99} \\
\end{tabular}
\end{ruledtabular}
\footnotetext[1]{1962 measurement with values given in cm$^{-1}$ and no error bars.}
\end{table}

To estimate the uncertainty in the RCC theory, we use the difference between the calculated and measured values of the magnetic-dipole coefficient $A$. From the difference of 3\%, we conservatively estimate the uncertainty in theory to be 5\%. The calculations yield $B/Q$ as $544.6$ MHz/b. Combined with our measured value of $B$, we get the quadrupole moment as $Q=2.46(12)$ b. This compares reasonably well with the 1962 value of $2.8(2)$ b \cite{ROM62}. The calculation also yields $C/\Omega$ as $-15.99$ kHz/(b$\times \mu_N$). From the measured value of $C$, we get the octupole moment as $\Omega=-34.4(21)$ b$\times \mu_N$.

\section{Conclusions}
In summary, we have made a precision measurement of hyperfine structure in the ${^3P}_2$ state of $^{173}$Yb, and see an unambiguous signature of the magnetic octupole coefficient $C$. The frequencies of the ${^3P}_2 \rightarrow {^3S}_1$ transition at 770 nm are measured using a Rb-stabilized ring-cavity resonator with an accuracy of 200 kHz. Second-order corrections are negligible at this level of precision, and do not affect the value of $C$. Using atomic-structure calculations for two-electron atoms \cite{MAA11}, we extract the nuclear {\em octupole moment} as $-34.4(21)$ b\,$\times \mu_N$, which is the first observation of this moment in two-electron atoms. We plan to complete similar measurements in the other odd isotope, $^{171}$Yb, and thereby get a handle on higher-order effects such as the Rosenthal-Breit effect and the Bohr-Weisskopf effect.

\begin{acknowledgments}
This work was supported by the Department of Science and
Technology. The computations were carried out on the 3 Teraflop cluster at Physical Research Laboratory. A.K.S.\ acknowledges financial support from the Council of Scientific and Industrial Research, India.
\end{acknowledgments}


%

\end{document}